\documentclass[fleqn,usenatbib]{mnras}

\usepackage{newtxtext,newtxmath}

\usepackage[T1]{fontenc}

\DeclareRobustCommand{\VAN}[3]{#2}
\let\VANthebibliography\thebibliography
\def\thebibliography{\DeclareRobustCommand{\VAN}[3]{##3}\VANthebibliography}

\usepackage{graphicx}	
\usepackage{amsmath}	

\title[Modeling of RR Lyrae stars with non-radial modes]{Detailed asteroseismic modeling of RR Lyrae stars with non-radial modes}

\author[H. Netzel et al.]{
H. Netzel,$^{1,2,3,5}$\thanks{E-mail: henia@netzel.pl}
L. Moln\'ar$^{1,2,4}$
and M. Joyce$^{1,2}$          
\\
$^{1}$Konkoly Observatory, Research Centre for Astronomy and Earth Sciences, E\"otv\"os Lor\'and Research Network; MTA Centre of Excellence (ELKH),\\ 
H-1121 Konkoly Thege Mikl\'os \'ut 15-17, Budapest, Hungary\\
$^{2}$MTA CSFK Lend\"ulet Near-Field Cosmology Research Group H-1121 Konkoly Thege Mikl\'os \'ut 15-17, Budapest, Hungary\\
$^{3}$ELTE E\"otv\"os Lor\'and University, Gothard Astrophysical Observatory, Szent Imre h. u. 112, 9700, Szombathely, Hungary\\
$^{4}$E\"otv\"os Lor\'and University, Institute of Physics, H-1117 P\'azm\'any P\'eter s\'et\'any 1/A, Budapest, Hungary\\
$^{5}$ Institute of Physics, Laboratory of Astrophysics, \'Ecole Polytechnique F\'ed\'erale de Lausanne (EPFL), Observatoire de Sauverny, 1290 Versoix, Switzerland
}

\date{Accepted XXX. Received YYY; in original form ZZZ}

\pubyear{2015}

\begin{document}
\label{firstpage}
\pagerange{\pageref{firstpage}--\pageref{lastpage}}
\maketitle

\begin{abstract}
Photometric observations from the last decade have revealed additional low-amplitude periodicities in many classical pulsators that are likely due to pulsations in non-radial modes. One group of multi-mode RR Lyrae stars, the so-called 0.61 stars, is particularly interesting. In these stars, the radial first overtone is accompanied by additional signals with period ratios around 0.61. The most promising explanation for these signals is pulsation in non-radial modes of degrees 8 and 9. 
If the theory behind the additional signals in the 0.61 stars is substantiated, it would allow us to use non-radial modes to study classical pulsators. We aim to perform asteroseismic modeling of selected 0.61 stars with independently determined physical parameters to test whether this assumption behind the modeling leads to correct results. Namely, we test whether the additional signals are indeed due to non-radial modes of the proposed moderate degrees.   
We selected a number of and RR Lyrae stars that are also 0.61 stars and have good observational constraints on their other physical parameters. We assume that the nature of those modes is correctly explained with non-radial modes of degrees 8 or 9. Using this assumption and observational constraints on physical parameters, we performed asteroseismic modeling to test whether the observed periods and period ratios can be reproduced. 
For the majority of selected targets, we obtained a good match between observed and calculated periods and period ratios. For a few targets however, the results obtained are ambiguous and not straightforward to interpret.
\end{abstract}




\section{Introduction}

RR Lyrae stars and classical Cepheids were considered to be very simple pulsators for many years. The majority of them pulsate in one or two radial modes. However, the last decade revolutionized our view on classical pulsators, with the help of large-scale sky surveys such as OGLE, and space-based missions such as {\it Kepler} and its continuation, K2, and TESS. When examined more closely, many of the classical pulsators show additional low-amplitude signals \citep[see a review by][and references therein]{smolec2021}. For many newly detected multi-mode groups among classical pulsators, the additional signals lack an explanation. However, one group is particularly interesting, because an explanation has been proposed for the nature of the additional signals. 

In many first-overtone RR Lyrae (or RRc type) stars, as well as first-overtone classical Cepheids, we detect additional signals with low amplitudes and short periods. Stars of this type form a first overtone--to--fundamental mode period ratio of about 0.61, henceforth referred to as 0.61 stars throughout this text. The first such stars were detected more than 10 years ago both among classical Cepheids \citep{soszynski2009} and RR Lyrae stars \citep{gruberbauer2007}. Many more such stars were detected since then, using both ground-based photometry \citep[see e.g.][and references therein]{jurcsik2015,smolec.sniegowska2016,netzel_census}, and space-based data \citep[see e.g.][and references therein]{moskalik2015,molnar2022}. A characteristic feature of the 0.61 stars is that in the Petersen diagram (the diagram of period ratio vs longer period), they form three sequences \citep[see e.g., Fig.~1 in][]{smolec2021}. Sometimes more than one additional signal is detected, in which case the star is placed onto more than one sequence in the Petersen diagram \cite[see fig.~8 in][]{netzel_census}. Signals at half-integer frequencies of the additional signal are also often detected. These are signals at $0.5f_{\rm X}$ or $1.5f_{\rm X}$, where $f_{\rm X}$ corresponds to the additional signal that forms a period ratio of about 0.61 with the first overtone.

An explanation for the additional signals was proposed by \cite{dziembowski2016}. In this scenario, the signal forming period ratio around 0.61 is the first harmonic of a non-radial mode. In the case of classical Cepheids, the observed sequences can be explained with non-radial modes of degrees $\ell=7$, 8 and 9 forming the top, middle and bottom sequences. In the case of RR Lyrae stars, the harmonics of non-radial modes of degrees $\ell=8$ and 9 correspond to the top and bottom sequences, respectively. The middle sequence observed in RR Lyrae stars is formed by the linear combination of two other modes. Due to the relatively high degree of the mode, it is subjected to cancellation effects; however, this affects the harmonic of the mode significantly less. Therefore it is typically easier to detect the harmonics that form the characteristic 0.61 period ratio than the mode frequencies. Still, we are sometimes able to observe the non-radial modes themselves, which are signals at $0.5f_{\rm X}$, where $f_{\rm X}$ is the frequency of the signal forming a period ratio between approximately 0.6 and 0.64. There are also a few detections of non-radial modes without their harmonics, but such instances are rare. A few such stars were reported by \cite{netzel_census} based on the analysis of a numerous sample of RRc stars in the OGLE Galactic fields. \cite{benko2021} reported such a detection in T~Sex. \cite{rajeev} found similar stars in the OGLE observations of classical Cepheids. 

Non-radial modes of different degrees are affected differently by the cancellation effect \citep[see Fig.~2 in][]{dziembowski2016}. Namely, the higher the degree of the mode, the stronger the cancellation effect. Also, modes of odd degrees are affected more strongly than modes of even degrees. In principle, it should be easier to detect $\ell = 8$ non-radial modes than $\ell = 7$ and 9 modes, according to the model by \cite{dziembowski2016}. The harmonic of the $\ell = 8$ corresponds to the top sequence in the case of RR Lyrae stars and to the middle sequence in the case of classical Cepheids. According to this prediction, signals at $0.5f_{\rm X}$ should correspond to signals from these sequences more frequently. \cite{netzel_census} showed that it is indeed the case for RR Lyrae stars and \cite{smolec.sniegowska2016} showed that for classical Cepheids. 

The possible explanation of these additional signals in classical pulsators through non-radial modes motivated the very first attempt to perform asteroseismic modeling of triple-mode 0.61 RR Lyrae stars \citep{netzel2022_rrlutnie}. They assumed mode identifications following \cite{dziembowski2016} and calculated a grid of theoretical models which was used to infer the physical parameters of the stars. This study was aimed at modeling the sample of 0.61 stars as a whole. The derived physical parameters were considered in the context of the whole sample and not for individual stars. In this study we take a step further in exploring the asteroseismic potential of classical pulsators with non-radial modes. We perform a detailed asteroseismic modeling of carefully selected targets which have good observational constraints on some of their physical parameters. We used these constraints in the modeling to explore whether the periods and period ratios can still be reproduced by the theoretical models that use the mode identification predicted by \cite{dziembowski2016}.

In Sec.~\ref{sec:method} we describe the selected targets and calculations of the models. The results of the modeling are presented in Sec.~\ref{sec:results} and discussed in Sec.~\ref{sec:discussion}. Sec.~\ref{sec:conclusions} contains conclusions.

\section{Method}\label{sec:method}

The targets had to fulfill several selection criteria. Naturally, they have to show the 0.61 signal. We note again that the 0.61 signal corresponds to the harmonic of the non-radial mode frequency: it was not required for stars to show the non-radial mode as well. During modeling we compared the period ratio that is formed by the harmonic of the non-radial modes with the first overtone.
As discussed by \cite{dziembowski2016}, the structures formed by the harmonic in the frequency spectra are narrower than those formed by the non-radial mode due to the non-linear interactions between the $2\ell +1$ multiplet components. Hence, the estimation of the frequency is more reliable when using the harmonic in the majority of the 0.61 stars. We used the frequency of the highest amplitude to compare with frequencies derived from theoretical models. 

The next selection criterion for the targets was the availability of independent and reliable physical parameters from spectroscopic observations: the effective temperature ($T_{\rm eff}$); and the [Fe/H] index, as a proxy for metallicity.

The last 
criterion was the availability of \textit{Gaia} astrometric data to obtain luminosity with relatively good accuracy \citep{gaia2016main}.

\begin{table}
    \centering
    \begin{tabular}{llll}
    ID & $M_G$ [mag] &$P_{\rm 1O}$ [d] & $P_{\rm S}/P_{\rm L}$  \\
    \hline
    CS Eri & $0.6200\pm0.0178$  &  0.3113307 & 0.61679 \\
    KIC\,8832417 & $1.264\pm 0.053$ &  0.2485464 & 0.61218  \\
    KIC\,5520878 & $0.932\pm 0.107$ &  0.2691699 & 0.63197   \\
    KIC\,4064484 & $0.484\pm 0.156$ &  0.3370019 & 0.61558  \\
    KIC\,9453114 & $0.432\pm 0.102$ &  0.3660809 & 0.61435  \\
    AE Boo  & $0.902\pm 0.039$ & 0.3149532 & 0.61316  \\
    AP Ser  & $0.523\pm 0.067$ & 0.3408494 & 0.61457  \\
    \end{tabular}
    \caption{Observed properties of selected stars. Consecutive columns provide ID of a star, absolute brightness in the $G$ band, observed first-overtone period, and period ratio between the first overtone and the additional signal.} 
    \label{tab:obs_params}
\end{table}

The selected targets are CS Eri, AE Boo, AP Ser, and four RRc stars from the original {\it Kepler} field: KIC\,4064484, KIC\,5520878, KIC\,8832417, and KIC\,9453114. The observed properties of \text{the} selected targets are presented in Table~\ref{tab:obs_params}. \text{From left to right, the} columns provide the name of the star, absolute brightness in the Gaia DR3 $G$ band, first-overtone period and period ratio with the additional signal, respectively.

\begin{figure}
    \centering
    \includegraphics[width=\columnwidth]{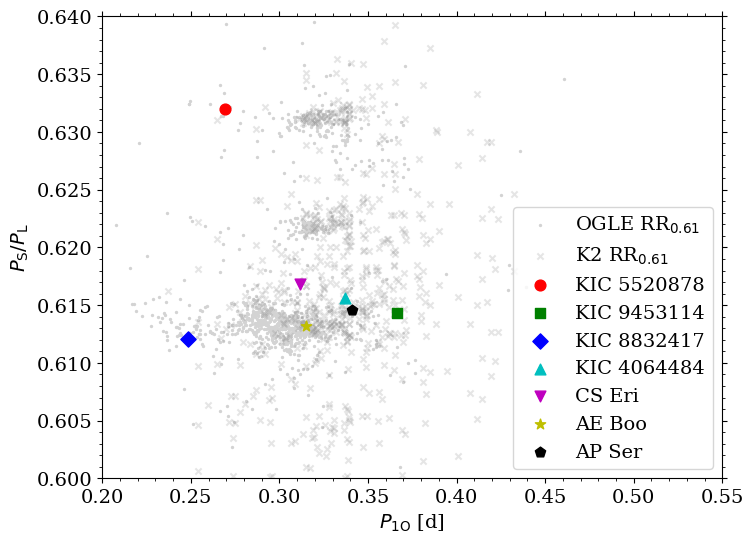}
    \caption{Petersen diagram displaying the selected targets, which are plotted with different colors and symbols as indicated in the legend. 0.61 stars from the Galactic bulge \protect\citep{netzel_census} and K2 Campaigns \protect\citep{netzel2023} are plotted with light-grey points and grey crosses, respectively, for reference.}
    \label{fig:pet_all}
\end{figure}

The positions of the selected targets in the Petersen diagram are presented in Fig.~\ref{fig:pet_all}. We included known 0.61 stars from the Galactic bulge and K2 Campaigns for reference \citep{netzel_census, netzel2023}. Six stars are located in the lower sequence, which corresponds to the harmonic of the $\ell=9$ non-radial mode according to the model of \cite{dziembowski2016}. The stars cover a wide range of first-overtone periods. The shortest first-overtone period in the studied sample is $P_{\rm 1O}=0.2485$\,d for KIC\,8832417, whereas the longest period is $P_{\rm 1O}=0.3661$\,d for KIC\,9453114. The typical period ratio of the lowest sequence based on the OGLE and K2 samples is around 0.613 \citep{netzel_census,netzel2023}. The most noticeable outlier from this value is CS Eri, for which the period ratio is around 0.61679. One star, KIC\,5520878, is a member of the upper-most sequence, which corresponds to the harmonic of the $\ell=8$ non-radial mode, according to the model of \cite{dziembowski2016}.

\begin{figure}
    \centering
    \includegraphics[width=\columnwidth]{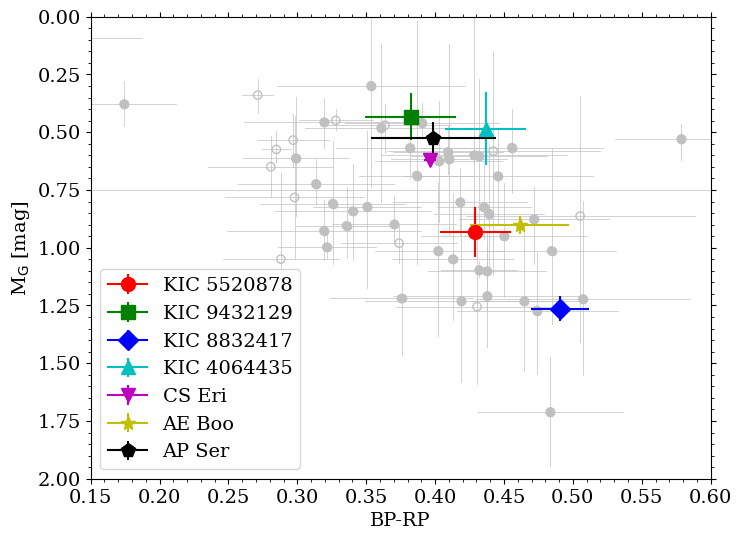}
    \caption{Color-magnitude diagram for selected targets. Each star is plotted with a different color and symbol. Grey points are bright RRc stars from the K2 sample \protect\citep{netzel2023}. Open symbols correspond to stars with no additional signals; filled -- to stars with additional signals. }
    \label{fig:cmd}
\end{figure}

We placed our sample on the color-magnitude diagram (CMD) in Fig.~\ref{fig:cmd} using absolute brightness in the \textit{Gaia} DR3 $G$ band and the $BP-RP$ color index (see Section \ref{sec:phys_param}). We also plotted the bright RRc stars from the K2 sample from \citet{netzel2023}, for reference. In the reference sample, filled symbols indicate detection of the additional signals, and open symbols indicate no detection of the additional signals. Stars analyzed in this work cover a similar range in absolute brightness but populate the redder side in color. We note that the spread observed in the absolute brightness is mostly due to the brightness dependence on metallicity \citep[see e.g.][]{plz_looijmans2023,clementini2023_gaia_rrl}. Stars studied here cover a wide range of metallicities, from the lowest metallicity of [Fe/H]=--2.13 measured for KIC\,9453114, to the highest metallicity of [Fe/H]=--0.18 measured for KIC\,520878 \citep{nemec2013}. The second highest metallicity of [Fe/H]=--0.27 is measured for KIC\,8832417, which also has the lowest observed brightness in the sample. Additional factors contributing to the observed spread in observed brightness for RR Lyrae stars are (1) different masses for individual objects and (2) how much they have evolved from the zero-age horizontal branch.

\subsection{Photometric observations}

Four RRc stars observed in the original {\it Kepler} field were analyzed by \cite{moskalik2015}. The 0.61 signal was detected in all of them. The additional signal is clearly apparent in frequency spectra after pre-whitening with the first overtone and its harmonics, as shown in Fig.~5 of \cite{moskalik2015}. We used the frequency values derived by \citealt{moskalik2015} (see their Table~4) in the present analysis.

For CS Eri, AE Boo and AP Ser, we performed frequency analysis using TESS data. CS Eri was observed during Sectors 3 and 30. AE Boo and AP Ser were observed during Sector 51. Light curves were extracted using the Lightkurve tool \citep{lightkurve} and converted to magnitudes using the TESS zero-point of $T_Z=20.44 \pm 0.05$\,mag (Vanderspek 2018, TESS Instrument Handbook). Derived periods and period ratios are included in Table~\ref{tab:obs_params}.

\begin{figure}
    \centering
    \includegraphics[width=\columnwidth]{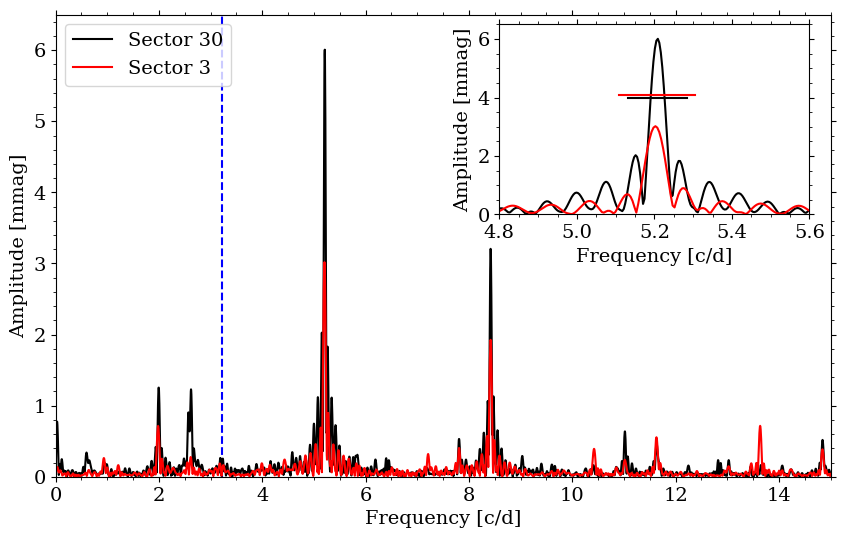}
    \caption{Frequency spectrum of CS Eri after prewhitening with the first overtone and its harmonics. The position of the first overtone is marked with the blue dashed line. Data from Sector 30 is plotted with black line, data from Sector 3 with red line. The meaning of the most significant peaks is indicated with labels. Zoom into the 0.61 signal is presented in the top right insert. The resolution of frequency spectrum is marked with horizontal lines.}
    \label{fig:cseri_spec}
\end{figure}

In the case of CS Eri, we analyzed both sectors separately. We plotted the frequency spectra in Fig.~\ref{fig:cseri_spec} after prewhitening with the first overtone and its harmonics. The position of the first overtone is marked with the blue dashed line. We plotted the frequency spectra from sectors 3 and 30 with red and black lines, respectively. The highest signal in both frequency spectra corresponds to the 0.61 signal. A more detailed view of this signal is presented in the top right insert in Fig.~\ref{fig:cseri_spec}, where we also marked the resolution of frequency spectra with horizontal solid lines (colors correspond to the respective Sectors). The 0.61 signal changed its amplitude from one sector to the other, but the frequency seems stable. Signals at $0.5f_{\rm X}$ and $1.5f_{\rm X}$ as well as combination signals are also well visible.

\begin{figure}
    \centering
    \includegraphics[width=\columnwidth]{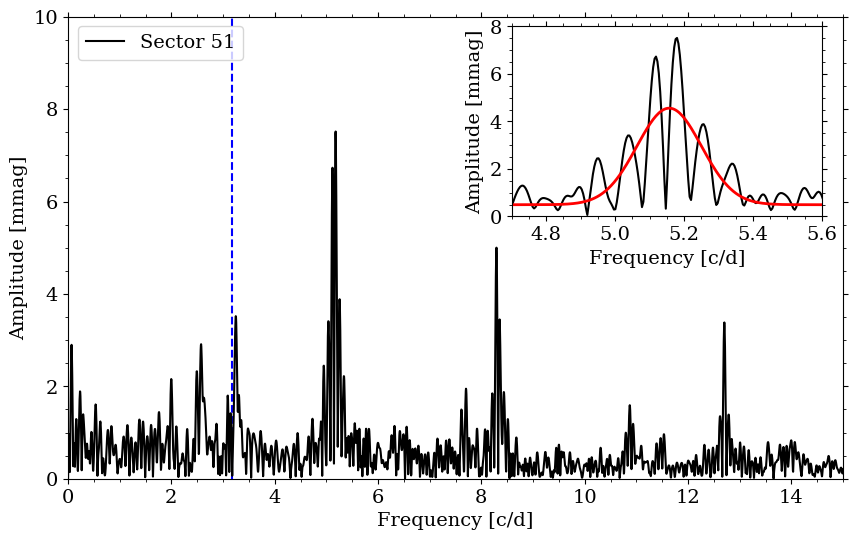}
    \caption{Frequency spectrum of AE Boo after prewhitening with the first overtone and its harmonics. The position of the first overtone is marked with the blue dashed line. The meaning of the most significant peaks is indicated with labels. Zoom into the 0.61 signal is presented in the top right insert. Red line corresponds to the Gaussian fit.}
    \label{fig:aeboo_spec}
\end{figure}

The frequency spectrum for AE Boo after prewhitening with the first overtone and its harmonics is presented in Fig.~\ref{fig:aeboo_spec}. The 0.61 signal is non-stationary, which manifests as the presence of close unresolved signals in frequency spectrum. We initially chose the frequency of the signal with the highest amplitude for further analysis. Using this value, the period ratio is 0.61316. There is no additional high-precision photometric data for AE Boo, therefore the non-stationarity of the signal cannot be investigated in more detail. Since the frequency of the highest-amplitude signal is not necessarily a good representation of the structure visible in frequency spectrum, we fitted a Gaussian function to the group of peaks. The fit is presented in Fig.~\ref{fig:aeboo_spec} with a red solid line. The centroid of the Gaussian function corresponds to the frequency of 5.15716284\,c/d, which results in the period ratio of 0.61566. Note, that in Table~\ref{tab:obs_params} we used period ratio based on the highest signal.

\begin{figure}
    \centering
    \includegraphics[width=\columnwidth]{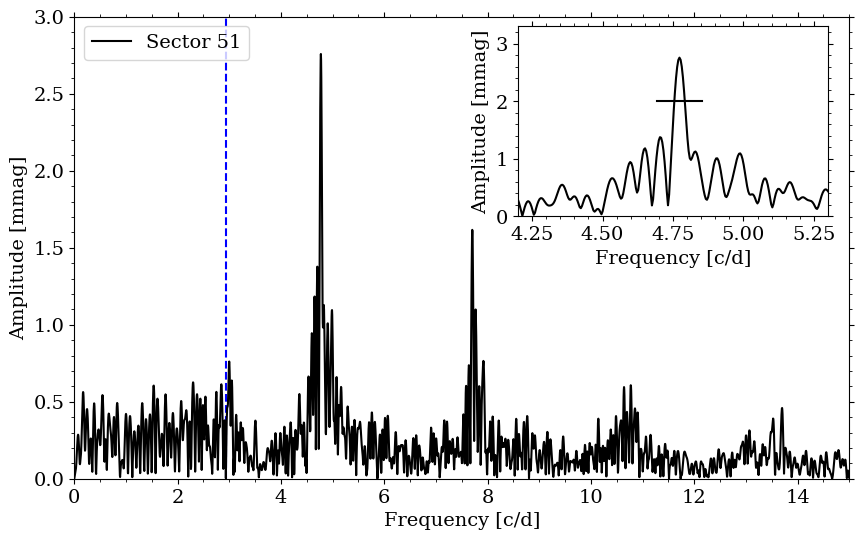}
    \caption{Frequency spectrum of AP Ser after prewhitening with the first overtone and its harmonics. The position of the first overtone is marked with the blue dashed line. The meaning of the most significant peaks is indicated with labels. Zoom into the 0.61 signal is presented in the top right insert. The resolution of frequency spectrum is marked with horizontal line.}
    \label{fig:apser_spec}
\end{figure}

The prewhitened frequency spectrum for AP Ser is presented in Fig.~\ref{fig:apser_spec}. Here, in contrast with AE Boo, the 0.61 signal is formed by a single peak. Further power excess is only slightly visible. Since this star was observed only during one Sector, we cannot investigate the temporal stability of the additional signal.

\subsection{Physical parameters}
\label{sec:phys_param}

Stars from the original {\it Kepler} field were observed spectroscopically by \cite{nemec2013}. We used their values for effective temperature, $T_{\rm eff}$, and metallicity proxy, [Fe/H] \citep[see Table~7 in][]{nemec2013}. 

\cite{crestani2021} derived various physical parameters for CS~Eri, AE Boo and AP Ser such as effective temperature, $T_{\rm eff}$, and the metallicity proxy, [Fe/H], based on spectroscopic observations (see their Table~2).

We then calculated the absolute \textit{G} brightness ($M_G$), and dereddened $BP-RP$ colors based on the Gaia DR3 main source catalog \citep{gaia2016main,gaia_dr3}, using the EDR3 distances calculated by \citet{edr3_distances}. We corrected for interstellar extinction using the \texttt{bayestar} 3D dust map and the \texttt{mwdust} software \citep{mwdust-2016,Green-2019}. For this step, we used the \texttt{Gaia} module of the \texttt{seismolab} python package \citep{bodi-seismolab}. We calculated bolometric brightnesses and luminosities using bolometric corrections based on \cite{creevey2022}.

\subsection{Theoretical models}

\begin{table*}
    \centering
    \begin{tabular}{lccccc}
    ID &  $M$ range [M$_\odot$] & $X$ range & $Z$ range & $\log T_{\rm eff}$ range & $\log L/L_\odot$ range \\
    \hline
    CS Eri  &  0.5--0.8 & 0.70--0.76 & 0.000099--0.0004 & 3.79--3.82 & 1.60--1.631 \\
    KIC\,8832417  & 0.5--0.8 & 0.70--0.76 & 0.005--0.019 & 3.835--3.855 & 1.325--1.385 \\
    KIC\,5520878 &   0.5--0.8 & 0.70--0.76 & 0.006--0.01 & 3.85--3.87 & 1.44--1.55  \\
    KIC\,4064484 &   0.5--0.8 & 0.70--0.76 & 0.00032--0.000735 & 3.80--3.823 & 1.61--1.76 \\
    KIC\,9453114 &   0.5--0.9 & 0.70--0.76 & 0.000076--0.00025 & 3.80--3.823 & 1.65--1.752 \\
    AE Boo &  0.5--0.8 & 0.70--0.76 & 0.00022--0.0009 & 3.81--3.832 & 1.479--1.525 \\
    AP Ser &  0.5--0.9 & 0.70--0.76 & 0.00024--0.0007 & 3.84--3.91 & 1.628--1.706 \\
    \end{tabular}
    \caption{Input parameters for the analysis. Consecutive columns provide ID of a star, observed first-overtone period and period ratio of the additional signal with the first overtone, and selected ranges of physical parameters: mass, hydrogen abundance, metal abundance, effective temperature and luminosity.}
    \label{tab:input_params}
\end{table*}

To calculate theoretical models, we used the Warsaw envelope pulsation code \citep{dziembowski1977}. First, the code constructs the model of an envelope \citep{paczynski1969} using the input parameters: mass, effective temperature, luminosity, hydrogen abundance ($X$), and metal abundance ($Z$). Then, the code uses the structure of an envelope to calculate periods and growth rates for radial first overtone and non-radial modes of degrees 8 and 9 \citep{dziembowski1977}.

Our calculations were performed in two steps. In the first step, we performed calculations of models to match the first-overtone period. For those calculations, we set accepted ranges of required physical parameters. The permitted ranges of input parameters for the analyzed stars are listed in Table~\ref{tab:input_params}. Consecutive columns provide name of a star, range of mass, range for hydrogen and metal abundance, and range for effective temperature and luminosity. Since there is no observational constraint for the mass, we chose a wide range of $M \in (0.5,0.8) M_\odot$ for most stars. However, for KIC\,9453114 and AP~Ser we extended the mass range to $0.9 M_\odot$. Typically, the mass of RR Lyrae stars is expected to be in a range of 0.5--0.8 M$_\odot$ based on evolutionary models \citep{marconi2015}. However, during the modeling carried out by \cite{netzel2022_rrlutnie}, the masses higher than 0.8 M$_\odot$ were obtained for stars having the longest first-overtone period. Therefore, we increased the allowed mass range for stars in this sample that have relatively long periods.

To obtain the estimate of hydrogen and metal abundance, we first transformed [Fe/H] to global metallicity [m/H] using Eq.~3 in \citet{Salaris-1993}:  

\begin{equation}
    {\rm[m/H]} \approx {\rm [Fe/H]} + \log(0.638\cdot 10^{\rm [\alpha/Fe]} + 0.362),
\end{equation}

where [$\alpha$/Fe] is the average $\alpha$--enhancement. We set [$\alpha$/Fe]\,=\,0.4.

The hydrogen and metal abundance was then calculated from [m/H] using the following formula:

\begin{equation}
    {\rm[m/H]}=\log(Z/{\rm Z_\odot})-\log(X/{\rm X_\odot}),
    \label{eq:mh}
\end{equation}

where $Z_\odot$ and $X_\odot$ are solar values of the metal and hydrogen abundance. We adopted $X_\odot=0.7381$ and $Z_\odot=0.0134$ from \cite{asplund2009}. 
We calculated the value of $Z$ for two values of $X$, namely for $X=0.7$ and $X=0.76$ and used them to set limits on the ranges. We note that for KIC\,5520878, the above approach would result in upper limit on $Z$ of above solar values. Therefore, we decided to artificially set the upper limit to $Z=0.01$. We also note that the slight inconsistency arises from the fact, that \cite{Salaris-1993} work predates the values of $X_\odot$ and $Z_\odot$ derived by \cite{asplund2009}. However, we do not expect this to significantly bias our results, since we adopted relatively wide ranges for $X$ and $Z$.

We used four values of the mixing-length parameter: 0.5, 1.0, 1.5 or 2.0. For each value we performed separate calculations. We used the OPAL opacities \citep{iglesias1996}.

In the first step, the goal was to find theoretical models which have a first-overtone period that matches the observed period. The physical parameters for the models were limited by observational constraints. To find the best-fitting models, we used the genetic algorithms method. We chose \texttt{geneticalgorithm} Python library\footnote{\url{https://github.com/rmsolgi/geneticalgorithm.git}}. 

\begin{figure}
    \centering
    \includegraphics[width=\columnwidth]{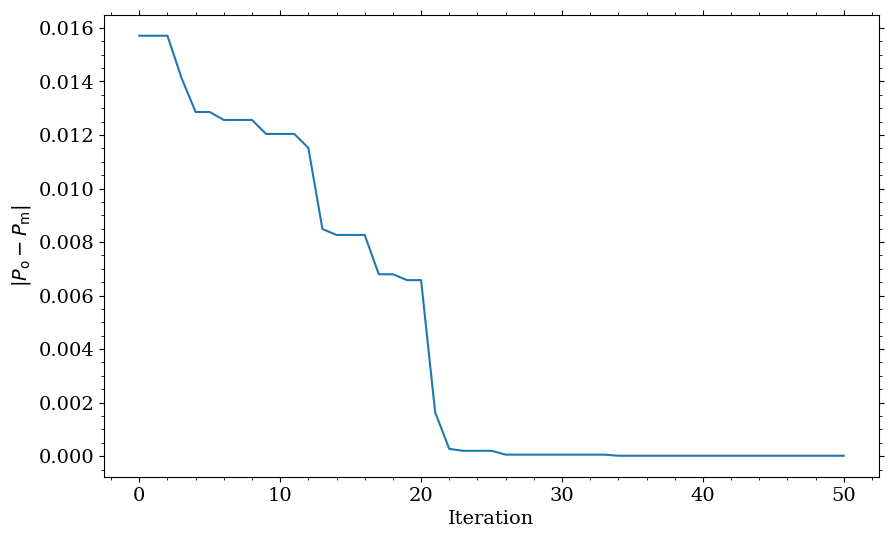}
    \caption{Example of a genetic algorithm optimization run on CS Eri. For each iteration, the lowest value of the absolute difference between the observed and calculated period is plotted.}
    \label{fig:ga_example}
\end{figure}

In this method, a chromosome represents the set of five input parameters (M, log\,L, log\,T$_{\rm eff}$, X, Z) and a fitness function that is an absolute difference between observed and calculated first-overtone period. We set the size of population to be 50 and limited a maximum number of iterations to 50. The rest of the parameters, such as mutation probability or crossover probability, remained default\footnote{The meaning of these parameters is explained in the README.md file in the \texttt{geneticalgorithm} library.}. An example of a run for a single star is presented in Fig.~\ref{fig:ga_example}, where the fitness function for each iteration is plotted. We note that we performed convergence testing. Namely, for individual stars we tested other values of parameters, including larger sizes of populations, higher numbers of iterations, and mutation probability, to test whether our results are robust.

In the second step, we used the physical parameters that correspond to models matching the first-overtone periods, to calculate period ratios formed by the harmonics of the non-radial modes. The pulsation code requires an estimated value of a dimensionless frequency of the non-radial mode. We used the same approach as in \cite{netzel2022_rrlutnie}. Namely, we estimated the dimensionless frequency, $\sigma$, from the fit to known 0.61 RR Lyrae stars \citep[see eq. 4 and 5 and fig. 1 in][]{netzel2022_rrlutnie}. We calculated 100 models in the range $\langle \sigma-0.1,\sigma+0.1 \rangle$ with a step of $\Delta \sigma=0.002$. Then, from 100 models we selected the one that has the highest driving rate for the non-radial mode. We note that typically there was more than one set of parameters that resulted in models fitting the first-overtone period with the same accuracy. We calculated period ratios formed by harmonics of non-radial modes for all such sets and chose those parameters that best reproduced the period ratio.

\begin{table*}
    \centering
    \begin{tabular}{llccccccccc}
    ID & $\alpha_{\rm MLT}$ & $P_{\rm 1O}$ [d] & $P_{\rm S}/P_{\rm L}$ & $\eta$ & $M$ & $\log T_{\rm eff}$ [K] & $\log L/L_\odot$  &  $X$ & $Z$  \\
    \hline
KIC\,5520878	&	2.0	&	0.2690296	&	0.63090	&	0.193	&	0.521	&	3.857	&	1.525	&	0.747	&	0.00943	\\
	&	1.5	&	0.2691535	&	0.63292	&	0.166	&	0.565	&	3.852	&	1.531	&	0.721	&	0.00920	\\
	&	\bf 1.0	&	\bf 0.2691535	&	\bf 0.63215	&	\bf 0.048	&	\bf 0.571	&	\bf 3.852	&	\bf 1.538	&	\bf 0.758	&	\bf 0.00694	\\
	&	0.5	&	0.2691535	&	0.63104	&	-0.144	&	0.523	&	3.861	&	1.546	&	0.736	&	0.00911	\\
	&		&	0.2691699	&	0.63197	&		&		&		&		&		&		\\
\hline																			
KIC\,9453114	&	2.0	&	0.3661002	&	0.61751	&	0.230	&	0.873	&	3.811	&	1.675	&	0.756	&	0.00014	\\
	&	1.5	&	0.3660159	&	0.61452	&	0.221	&	0.778	&	3.815	&	1.658	&	0.740	&	0.00020	\\
	&	1.0	&	0.3661002	&	0.61478	&	0.197	&	0.828	&	3.819	&	1.697	&	0.726	&	0.00008	\\
	&	\bf 0.5	&	\bf 0.3661002	&	\bf 0.61428	&	\bf 0.222	&	\bf 0.789	&	\bf 3.814	&	\bf 1.663	&	\bf 0.726	&	\bf 0.00015	\\
	&		&	0.3660809	&	0.61435	&		&		&		&		&		&		\\
\hline																			
KIC\,8832417	&	2.0	&	0.2485421	&	0.61121	&	0.323	&	0.501	&	3.836	&	1.378	&	0.754	&	0.01387	\\
	&	1.5	&	0.2483133	&	0.61089	&	0.319	&	0.502	&	3.836	&	1.382	&	0.751	&	0.01163	\\
	&	1.0	&	0.2470017	&	0.61135	&	0.299	&	0.507	&	3.836	&	1.384	&	0.727	&	0.00504	\\
	&	\bf 0.5	&	\bf 0.2482561	&	\bf 0.61161	&	\bf 0.325	&	\bf 0.505	&	\bf 3.836	&	\bf 1.384	&	\bf 0.757	&	\bf 0.01848	\\
	&		&	0.2485464	&	0.61210	&		&		&		&		&		&		\\
\hline																			
KIC\,4064484	&	2.0	&	0.3369768	&	0.61521	&	0.258	&	0.782	&	3.819	&	1.627	&	0.751	&	0.00066	\\
	&	\bf 1.5	&	\bf 0.3369768	&	\bf 0.61554	&	\bf 0.191	&	\bf 0.849	&	\bf 3.810	&	\bf 1.620	&	\bf 0.754	&	\bf 0.00043	\\
	&	1.0	&	0.3366666	&	0.61463	&	0.186	&	0.880	&	3.816	&	1.655	&	0.747	&	0.00072	\\
	&	0.5	&	0.3369768	&	0.61471	&	0.206	&	0.798	&	3.822	&	1.654	&	0.704	&	0.00041	\\
	&		&	0.3370019	&	0.61558	&		&		&		&		&		&		\\
\hline																			
CS\,Eri	&	2.0	&	0.3204793	&	0.61696	&	0.211	&	0.783	&	3.818	&	1.601	&	0.749	&	0.00011	\\
	&	1.5	&	0.3128240	&	0.61572	&	0.160	&	0.795	&	3.820	&	1.600	&	0.722	&	0.00017	\\
	&	1.0	&	0.3123921	&	0.61558	&	0.150	&	0.795	&	3.819	&	1.601	&	0.702	&	0.00012	\\
	&	\bf 0.5	&	\bf 0.3113150	&	\bf 0.61548	&	\bf 0.158	&	\bf 0.795	&	\bf 3.820	&	\bf 1.602	&	\bf 0.709	&	\bf 0.00010	\\
	&		&	0.3113307	&	0.61679	&		&		&		&		&		&		\\
\hline																			
AE\,Boo	&	2.0	&	0.3149198	&	0.61109	&	0.304	&	0.550	&	3.827	&	1.509	&	0.719	&	0.00083	\\
	&	1.5	&	0.3165920	&	0.60984	&	0.313	&	0.505	&	3.826	&	1.480	&	0.757	&	0.00050	\\
	&	\bf 1.0	&	\bf 0.3149198	&	\bf 0.61307	&	\bf 0.231	&	\bf 0.668	&	\bf 3.812	&	\bf 1.512	&	\bf 0.740	&	\bf 0.00067	\\
	&	0.5	&	0.3149198	&	0.61349	&	0.235	&	0.673	&	3.810	&	1.512	&	0.730	&	0.00056	\\
	&		&	0.3149451	&	0.61316	&		&		&		&		&		&		\\
\hline																			
AP\,Ser	&	2.0	&	0.3408788	&	0.61107	&	0.307	&	0.567	&	3.849	&	1.656	&	0.746	&	0.00039	\\
	&	1.5	&	0.3415073	&	0.61232	&	0.242	&	0.651	&	3.850	&	1.705	&	0.713	&	0.00068	\\
	&	1.0	&	0.3408788	&	0.61093	&	-0.124	&	0.519	&	3.864	&	1.691	&	0.743	&	0.00030	\\
	&	\bf 0.5	&	\bf 0.3408788	&	\bf 0.61295	&	\bf 0.081	&	\bf 0.669	&	\bf 3.845	&	\bf 1.695	&	\bf 0.722	&	\bf 0.00057	\\
	&		&	0.3408494	&	0.61457	&		&		&		&		&		&		\\
    \end{tabular}
    \caption{Results of modeling of selected targets. There five rows for each star. First to fourth rows correspond to the results for $\alpha_{\rm MLT}=2.0$, 1.5, 1.0 and 0.5, respectively. For reference, in the fifth row we provided observed values of period and period ratio. For the first four rows for each star, the consecutive columns provide ID of the star, $\alpha_{\rm MLT}$, calculated first-overtone period and period ratio, growth rate $\eta$, mass, logarithm of effective temperature and luminosity, hydrogen and metal abundance. We marked the best fitting solutions with bold font.}
    \label{tab:results}
\end{table*}

We repeated the described two steps of calculations for four values of $\alpha_{\rm MLT}$. Note, that $\alpha_{\rm MLT}$ was not optimized in the genetic algorithm runs. We included in the chromosome vector only the parameters which are observationally contrained. 


\begin{figure*}
    \centering
    \includegraphics[width=\columnwidth]{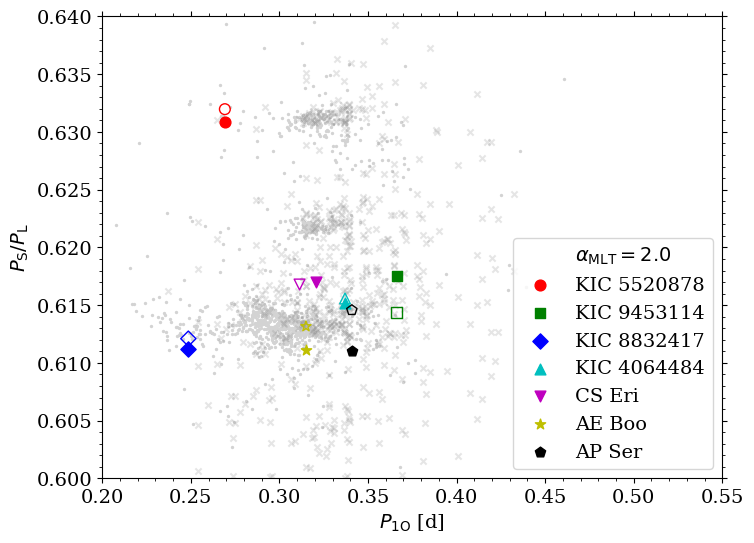}
    \includegraphics[width=\columnwidth]{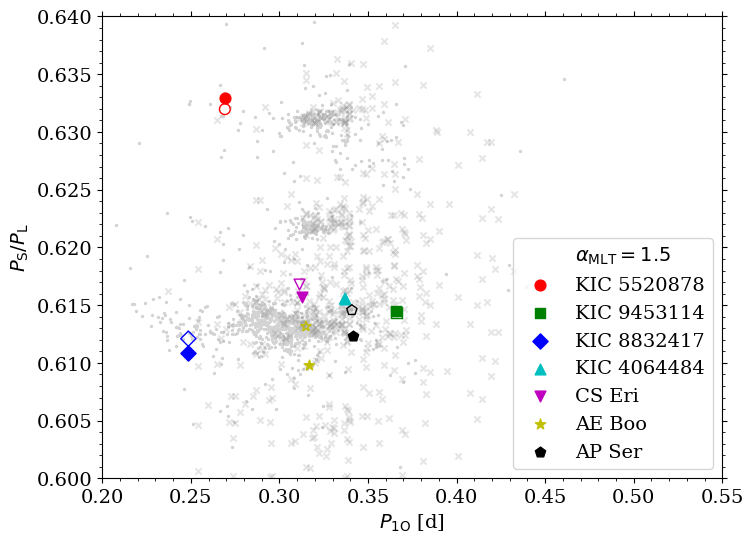}
    \includegraphics[width=\columnwidth]{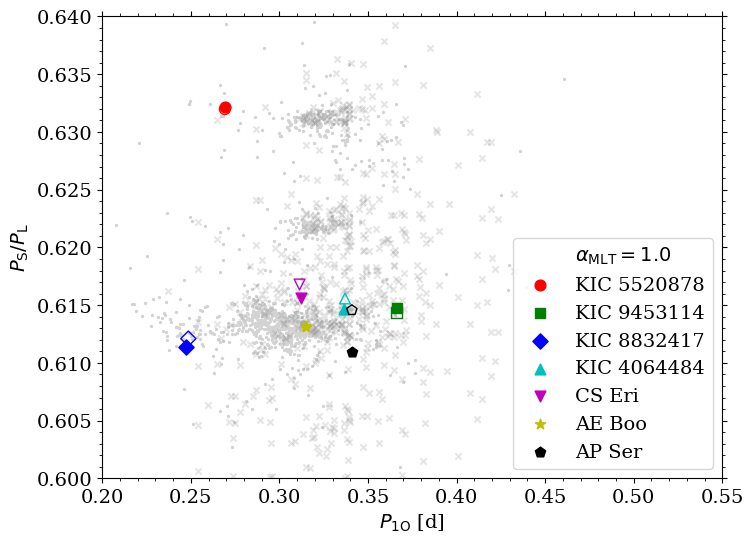}
    \includegraphics[width=\columnwidth]{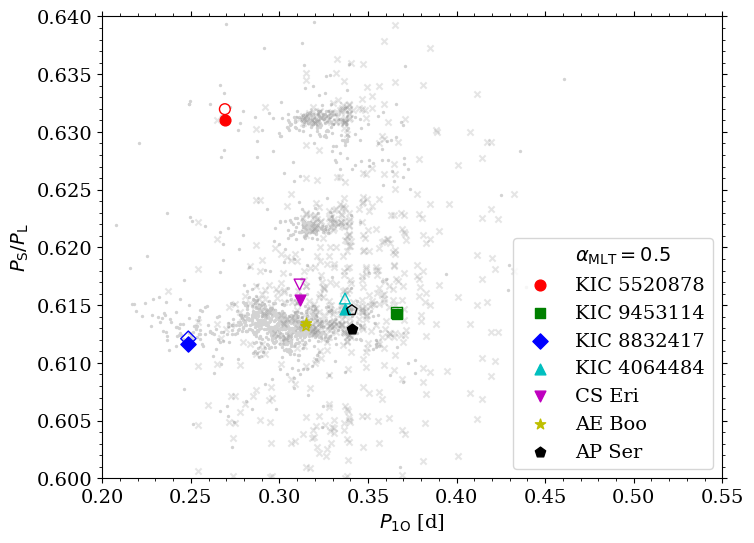}
    \caption{Results presented in the form of the Petersen diagram. Top left panel: results for $\alpha_{MLT}=2.0$. Top right panel: results for $\alpha_{MLT}=1.5$. Bottom left panel: results for $\alpha_{MLT}=1.0$. Bottom right panel: results for $\alpha_{MLT}=0.5$. Observed values of periods and period ratios are plotted with open symbols. Calculated values are plotted with filled symbols. For a reference, we included RR$_{0.61}$ stars from the OGLE and K2 samples, same as in Fig.~\ref{fig:pet_all}.}
    \label{fig:pet_all_results}
\end{figure*}

\section{Results}\label{sec:results}

The results for the selected targets are collected in Table~\ref{tab:results}. Consecutive columns provide the ID of the star, $\alpha_{\rm MLT}$, first-overtone period, period ratio, growth rate of the non-radial mode, mass, effective temperature, luminosity, hydrogen and metal abundances. For each star there are five rows. The first four rows show the results for different values of $\alpha_{\rm MLT}$. In the last row we provided observed values of periods and period ratios for reference. For all seven stars we were able to obtain at least one model which reproduced the first-overtone period and has a linearly unstable non-radial mode. Only in the case of KIC\,5520878 and $\alpha_{\rm MLT}=0.5$, as well as AP\,Ser and $\alpha_{\rm MLT}=1.0$ the calculations resulted in linearly stable non-radial modes.

For some stars we obtained better match of period ratios for $\alpha_{\rm MLT}=0.5$. This is the case for KIC~9453114, KIC~8832417, and AP Ser. In the case of KIC\,5520878 and AE\,Boo, the better match was obtained using $\alpha_{\rm MLT}=1.0$. The calculations with $\alpha_{\rm MLT}=1.5$ resulted in the best-fitting model only in the case of KIC\,4064484. Interestingly, in the case of CS Eri, the fit to the first-overtone period is better for $\alpha_{\rm MLT}=0.5$, while the match between observed and calculated period ratio is better for $\alpha_{\rm MLT}=2.0$. For further discussion of the results, in the case of CS Eri, we used the model corresponding to $\alpha_{\rm MLT}=0.5$, as it provided the overall better fit for both period and period ratio. It is noteworthy that no star is best fit by a model with $\alpha_{\rm MLT}=2.0$ (see discussion in Sec.~\ref{sec:discussion}).

Calculated periods and period ratios for all targets are plotted in Fig.~\ref{fig:pet_all_results}. Different panels correspond to four values of $\alpha_{\rm MLT}$. Each star is plotted with different colors and symbols, as indicated in the key. Open symbols correspond to observed values of periods and period ratios. Filled symbols correspond to theoretical values from the best-fitting model. Note that we plotted results for all stars, regardless of whether the non-radial mode is linearly unstable in the models. In the case of AE Boo, a significant improvement in reproducing the observed period ratio is visible for lower values of $\alpha_{\rm MLT}$. Interestingly, in the case of AP Ser, all values of $\alpha_{\rm MLT}$ result in unsatisfactory model fits.

\begin{figure}
    \centering
    \includegraphics[width=0.5\textwidth]{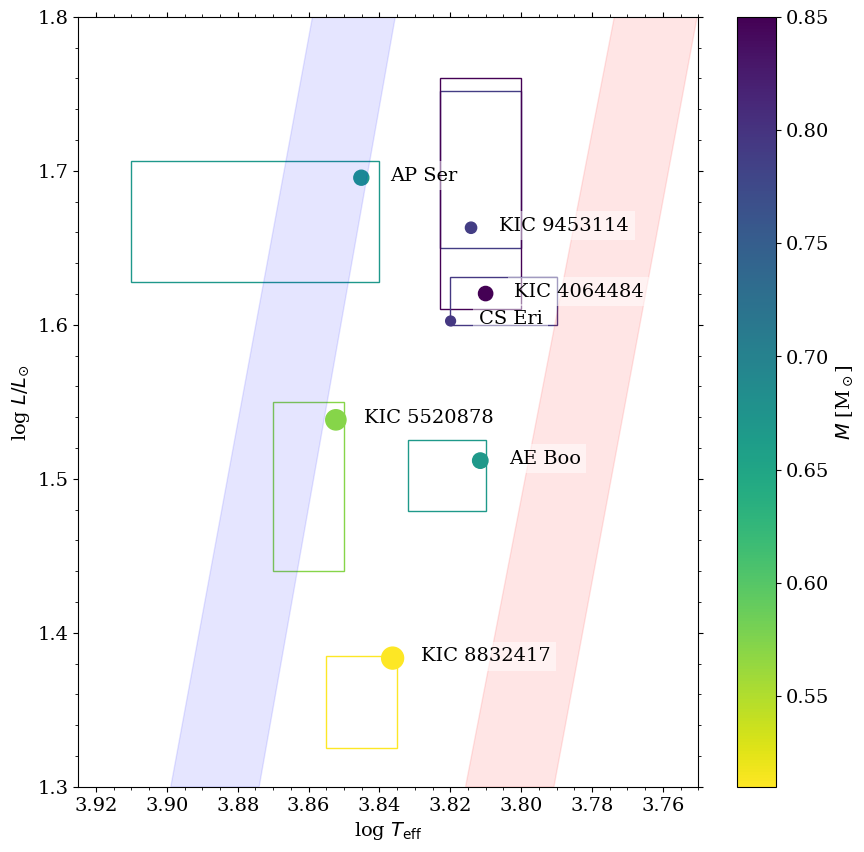}
    \caption{Results presented in the Hertzsprung-Russell diagram. Each star was plotted using the calculated effective temperatures and luminosities. The ranges of effective temperature and luminosity based on observations are indicated by the rectangles. Size of the points correspond to calculated metal abundance, Z. Calculated mass is color-coded as indicated in the key.  We marked the positions of the blue and red edges of the instability strip based on \protect\cite{marconi2015} with blue and red stripes for a range of metallicities that we obtained for the studied stars.}
    \label{fig:hr_results}
\end{figure}

In Fig.~\ref{fig:hr_results} we plotted stars for which we obtained models with non-radial modes in the form of the Hertzsprung-Russell (HR) diagram. The majority of stars are located in the central parts of the instability strip. Two stars are located close to the blue strip that represents the range of locations for the blue edge. These are AP Ser and KIC\,5520878. The distribution of stars in the HR diagram (Fig.~\ref{fig:hr_results}) matches the distribution of stars in the CMD in Fig.~\ref{fig:cmd} closely.

Two trends are visible in the HR diagram. Metal abundance, indicated by the size of the symbols, tends to decrease with increasing luminosity. The second trend (if we exclude AP Ser) is that the mass increases with luminosity. We note however, that the fit for AP Ser is not satisfactory (see Table~\ref{tab:results}). 

We did not compare the results of our pulsation analysis with evolutionary tracks, because such comparison is not straightforward. Namely, the available evolutionary tracks are calculated for the solar values of $\alpha_{MLT}$, and not for the values considered in this study. Therefore, it would lead to significant differences in tracks \cite[see Fig.~3 and a discussion in][]{joyce.tayar2023}, and would impede our efforts to draw reliable conclusions from comparing of our results to publicly available evolutionary tracks.

\subsection{Results on individual stars}

\paragraph*{KIC 5520878}
This is the only star that is the member of the top sequence, i.e., the additional signal corresponds to the non-radial mode degree $\ell=8$ according to \cite{dziembowski2016}. The best-fitting model corresponds to $\alpha_{\rm MLT}=1.0$. The relative error of period ratio is 0.03\%. Interestingly, the model calculated with $\alpha_{\rm MLT}=0.5$ has a linearly stable non-radial mode. It is located in the blue part of the instability strip on the HR diagram in Fig.~\ref{fig:hr_results}. 

\paragraph*{KIC 9453114}
The best-fitting model corresponds to $\alpha_{\rm MLT}=0.5$. Period ratio is very well reproduced. The relative error of calculated period ratio is only 0.01\%. The theoretical mass is high, 0.789\,$M_\odot$, but is still within what is expected from RR Lyrae stars. KIC\,9453114 is located in the center of the instability strip. The effective temperature of the model is around the middle of the allowed range. However, the obtained luminosity falls within the lower limit of the acceptable range. 

\paragraph*{KIC 8832417}
The best-fitting model corresponds to $\alpha_{\rm MLT}=0.5$. Period ratio is reproduced with a relative error of 0.08\%. Calculated mass is low, 0.505\,$M_\odot$, which is close to the lower limit of the allowed mass range. KIC\,8832417 is the lowest-mass, lowest-luminosity and highest-metallicity star in our sample. Interestingly, both the effective temperature and the luminosity obtained from the modeling are almost at the limits of the allowed ranges. In particular, the luminosity is at the high limit of values allowed from observations.

\paragraph*{KIC 4064484}
This is the only star in our sample where the best-fitting model corresponds to $\alpha_{\rm MLT}=1.5$. Period ratio is reproduced with a relative error of 0.006\%. Given the relatively long first-overtone period of 0.337\,d for KIC,4064484, the upper limit on the mass used for modeling has been increased to 0.9\,M$_\odot$. The obtained mass is 0.849\,M$_\odot$. On the HR diagram, KIC\,4064484 is located in the center of the instability strip. The allowed range for luminosity was relatively large, and the derived value is close to its lower limit. 

\paragraph*{CS Eri}
The best-fitting model corresponds to $\alpha_{\rm MLT}=0.5$. The period ratio is reproduced with a relative error of 0.2\%, which is a significantly worse result than for the {\it Kepler} stars. However, CS Eri was observed during two TESS Sectors with a short timebase of observations of only 20 and 26 days per Sectors 3 and 30, respectively. This timebase corresponds to frequency resolution of 0.05\,d${-1}$ and 0.038\,d${-1}$, respectively, which translates to an uncertainty of of 0.0111 and 0.0086 in the period ratios for Sector 3 and 30. On the other hand, frequency spectra for both Sectors suggest that the additional signal is relatively stable and its frequency can be reliably estimated (see Fig.~\ref{fig:cseri_spec}). Interestingly, the derived values of effective temperature and luminosity are at the limits of ranges allowed based on observations.

\paragraph*{AE Boo}
AE Boo was only observed during one TESS Sector for a duration of 24.58\,d. Without longer photometric observations we cannot test whether the frequency and amplitude of the additional signal change in time. The frequency spectrum (Fig.~\ref{fig:aeboo_spec}) suggests that this might be the case for AE Boo. On top of that, the uncertainty of period ratio due to short timebase of observation is 0.009. 

Without additional observations we cannot unambiguously determine the period ratio, hence the results are less certain for AE Boo than for the {\it Kepler} stars or CS Eri. As a first approximation we used the period ratio determined from the frequency of the highest peak in the additional signal group. This resulted in an observed period ratio of 0.61316, which is reproduced by the model with $\alpha_{\rm MLT}=1.0$ with relative error of 0.01\%. Interestingly, for high values of $\alpha_{\rm MLT}$, there is a significant difference in observed and calculated period ratios (see Fig.~\ref{fig:pet_all_results}).

Additionally, we checked the results using the centroid of the Gaussian fit is instead of the highest-amplitude signal (see Fig.~\ref{fig:aeboo_spec}). Using value of the centroid results in the period ratio of 0.61566. This period ratio was not reproduced by the models.

In the HR diagram, AE Boo is located in the red part of the instability strip. The calculated effective temperature is also close to the low temperature limit set by the observations.

\paragraph*{AP Ser}
AP Ser faces the same issues related to the reliable determination of the period ratio as AE Boo. Namely, analysis of only the one available sector of TESS observations results in an uncertainty of the period ratio of 0.01. Without a sufficiently long timebase of photometric observations we cannot trace amplitude and frequency variations of the additional signal. 

The modeling of AP Ser was unsuccessful. For all values of $\alpha_{\rm MLT}$, the calculated period ratio is too low with respect to the observed value. In the best-fitting model calculated for $\alpha_{\rm MLT}=1.0$ the non-radial mode is linearly stable. For the three other values of $\alpha_{\rm MLT}$, the calculated period ratio has a relative error from 0.3\% to 0.6\%.

On the HR diagram, AP Ser is located close to the blue edge of the instability strip. It is worth noting that the allowed ranges of parameters tend to favor higher effective temperatures, which extend beyond the theoretically predicted position of the instability strip. Effective temperature for the best-fitted model is located in the low-temperature part of the allowed range. AP Ser is also the most luminous star from the studied sample.

\section{Discussion}\label{sec:discussion}

For most stars, we obtained a good fit between observed and calculated periods and periods ratios (see the Petersen diagram in Fig.~\ref{fig:pet_all_results}). Only for one star, AP Ser, did we not find a well fitted model. However, AP Ser is one of the two stars that has only one sector of TESS observations available. The additional signals in 0.61 stars can have strong temporal variations in amplitude and/or phase, resulting in wide structures in the frequency spectra. Therefore, frequency estimation can be uncertain when only short amounts of photometry are available. In contrast, the four RRc stars from the {\it Kepler} field, have long time base of observations which allowed us to trace the variability of additional signals well \citep[see fig. 6 and 7 in][]{moskalik2015}. Subsequently, the representative frequency for additional signal can be reliably estimated from the frequency spectra. On the other hand, stars from the {\it Kepler} field are relatively faint, which results in larger uncertainty in distance estimation than in the case of the other three targets from our sample. Consequently, the uncertainty in luminosity is the highest for stars from the {\it Kepler} field.

For A/F-type stars, the stellar interior hosts a nested convective structure: there is a convective core at the center, which is surrounded concentrically by a radiative zone and a thin, outer convective envelope. The thinner the convective envelope, the less the choice of $\alpha_{\rm MLT}$ matters in stellar evolution calculations: this is because the mixing length theory of convection parameterizes an entropy jump between the innermost and outermost extents of the convection zone. When the difference between the temperature gradient at the top of the convection zone and the temperature gradient at the lower, interior edge of the convection zone becomes small, as is the case for stars with thin convective envelopes, the impact of changing the convective efficiency parameter becomes negligible \citep[see section 6 of][]{joyce.tayar2023}. As such, it is not an indication of inconsistency that very different values of $\alpha_{\rm MLT}$ are found to best fit different stars in our sample. However, it is also true that thin convective envelopes are unlikely to be well fit by large values of $\alpha_\text{MLT}$ (i.e., values that indicate high convective efficiency), because the maximal extent of the convective mean-free path is physically limited. This is supported by the fact that no star in the sample prefers the highest value of $\alpha_\text{MLT}$ considered.

One of the most important parameters in the case of RR~Lyrae stars is the mass. A direct determination of the mass of an RR~Lyrae star is not possible, as there are no known eclipsing binary systems that contain an RR~Lyrae star. The only good candidate turned out to be an RR~Lyrae impostor \citep{pietrzynski2012}. Therefore, the only way to obtain estimation of the masses of RR~Lyrae stars is through indirect methods. For instance, \cite{molnar2015} used hydrodynamic modeling of RRd stars and was able to derive the masses. 

\cite{simon.clement1993} developed formulae to predict physical parameters from the light curve shape based on hydrodynamic modeling of RRc stars and light curve fitting of stars in globular clusters (see their Eq.~2, 3, 4, and 5). We used periods and Fourier phase differences, $\phi_{31}$, for CS Eri, AP Ser and AE Boo. For helium abundance, we used four different values: $Y=0.30$, 0.25, 0.20, and value calculated from the best fitting model. We obtained masses and luminosities, which are in disagreement with results from our modeling. The formulae based on light curve shapes predict mass of around 1.0 M$_{\odot}$ for CS Eri, 0.73 M$_{\odot}$ for AE Boo, and 0.80 M$_{\odot}$ for AP Ser. The predicted masses are higher than the masses obtained in this work. The lowest discrepancy is for AE Boo, while the highest discrepancy is for CS Eri. The predicted luminosities are around 86 to 120 L$_{\odot}$, which is also higher than luminosities calculated in the present analysis.

\cite{marsakov2019} estimated masses of RR Lyrae stars in globular clusters based on theoretical evolutionary tracks. In particular, for CS Eri they estimated a mass of 0.63 M$_{\odot}$, effective temperature of $T_{\rm eff}=6750$ K, metallicity of [Fe/H]\,=\,--1.70 and alpha element enhancement of [$\alpha$/Fe]\,=\,0.35. The mass that we obtained with modeling is higher, 0.795 M$_{\odot}$. However, the effective temperature that was obtained by \cite{marsakov2019} is outside of the effective temperature range used for modeling  based on \cite{crestani2021}. Moreover, the metallicity estimated by \cite{crestani2021} is also lower than the value given by \cite{marsakov2019}. 

\begin{figure}
    \centering
    \includegraphics[width=\columnwidth]{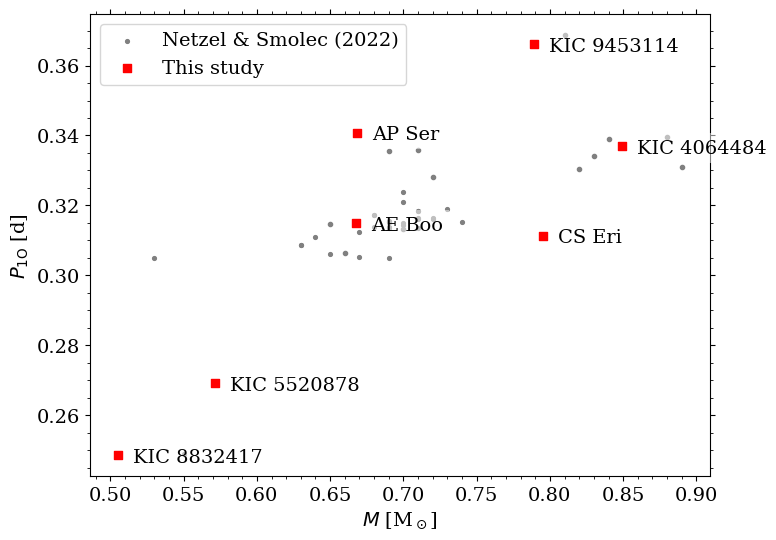}
    \caption{The relation between the observed first-overtone period and calculated mass. Stars analyzed here are plotted with red squares. For reference, we plotted with gray points RRc stars with non-radial modes modeled by \protect\cite{netzel2022_rrlutnie}.}
    \label{fig:mp1o}
\end{figure}

\cite{netzel2022_rrlutnie} used at least triple-mode RR~Lyrae stars pulsating with additional non-radial modes to estimate the physical parameters based on asteroseismic modeling.  
In Fig.~\ref{fig:mp1o} we compared our results on calculated masses with results for RRc stars from \cite{netzel2022_rrlutnie}. We obtained similar results to Fig.~9 in their paper: namely, the mass increases with increasing first-overtone period. In particular, for a group of long-period stars, the masses derived by \cite{netzel2022_rrlutnie} were above 0.8\,M$_\odot$. This was the case for RRd stars with non-radial mode of degree $\ell=9$ or some of RRc stars from the K2 observations. Unfortunately, we do not have RRd stars in our sample. 

In the case of K2 RRc stars from \cite{netzel2022_rrlutnie}, the short time-base of observations might affect the reliability of estimating the representative frequency of the additional signal and consequently it might affect the modeling results. In our sample there are stars with relatively long first-overtone period: AP Ser (0.34\,d), KIC\,4064484 (0.337\,d), and KIC\,9453114 (0.37\,d). For AP Ser, we did not obtain a well fitted model. In the case of the two stars from the {\it Kepler} field, the long time of observations makes the estimation of the frequency reliable. Indeed for those two stars we obtained high mass of around 0.8\,M$_\odot$ or higher.

\section{Conclusions}\label{sec:conclusions}
We preformed a detailed modeling of selected RR Lyrae stars that show the so-called 0.61 signals. For the modeling, we chose RR Lyrae stars that have their metallicities and effective temperatures constrained via spectroscopic observations, and their luminosities constrained by Gaia data. Our input sample consisted of 7 RRc stars. Four stars were from the original {\it Kepler} field and three stars were observed by the TESS mission. All stars but one showed the additional signal that, according to the theory of \cite{dziembowski2016}, corresponds to the non-radial mode of degree $\ell=9$. Only in one star, KIC\,5520878, would the additional signal correspond to the non-radial mode of degree $\ell=8$.  We calculated models using the Warsaw envelope code with the aforementioned observational constraints to reproduce first-overtone period. For models that reproduced first-overtone period, we calculated periods and growth rates of non-radial modes of degrees 8 or 9. Finally, we compared theoretical period ratios that are formed by the first harmonics of non-radial modes of degrees $\ell=8$ or 9 and the first overtone, with the observed values of period ratios formed by the additional signals and the first overtone.

From the input sample containing seven stars we were able to obtain satisfactory fits for six of them within the observational constraints of the physical parameters. Only in the case of AP Ser, does the best-fitted model not reproduce the observed period and period ratio well. However, AP Ser lacks sufficiently long timebase of observations to reliably determine the observed period ratio. Longer monitoring of AP Ser would allow us to investigate the discrepancy between observed and modeled values in more detail. 

The masses obtained from the modeling cover a wide range of 0.50--0.90\,M$_\odot$. Interestingly, we observe a trend within the sample: the longer the period of the first overtone, the higher the mass of the best-fitted model. 

In the modeling, we adopted four values of $\alpha_{\rm MLT}$. For three stars, KIC\,9453114, KIC\,8832417, and CS Eri, the best-fitting models correspond to $\alpha_{\rm MLT}=0.5$. For two stars, KIC\,5520878, and AE Boo, we obtained the best fit using $\alpha_{\rm MLT}=1.0$. The best-fitting model for KIC\,4064484 was calculated using $\alpha_{\rm MLT}=1.5$. The models calculated with $\alpha_{\rm MLT}=2.0$ were not selected in any instance.

This work was aimed at testing whether the theory developed by \cite{dziembowski2016} regarding the nature of the additional 0.61 signals in RR Lyrae stars can lead to reliable results in theoretical modeling when confronted with independently determined observational constraints on physical parameters. For the majority of stars, the results are reasonable and in agreement with the observed physical parameters. We therefore gained another argument in favor of the theory explaining the nature of the observed additional signals forming period ratio of around 0.61 in classical pulsators \citep{dziembowski2016}. Finally, one of the significant outcomes is that we demonstrated a new way to determine seismic masses of RR Lyrae stars.

\section*{Acknowledgments}

H.N. has been supported by the \'UNKP-22-4 New National Excellence Program of the Ministry for Culture and Innovation from the source of the National Research, Development and Innovation Fund. 
This project has been supported by the Lend\"ulet Program of the Hungarian Academy of Sciences, project No. LP2018-7/2020, by the `SeismoLab' KKP-137523 \'Elvonal and  NN-129075 grants of the Hungarian Research, Development and Innovation Office (NKFIH). 
M.J. gratefully acknowledges funding of MATISSE: \textit{Measuring Ages Through Isochrones, Seismology, and Stellar Evolution}, awarded through the European Commission's Widening Fellowship. This project has received funding from the European Union's Horizon 2020 research and innovation program.
This work uses frequency analysis software written by R. Smolec. 
This work has made use of data from the European Space Agency (ESA) mission {\it Gaia} (\url{https://www.cosmos.esa.int/gaia}), processed by the {\it Gaia} Data Processing and Analysis Consortium (DPAC, \url{https://www.cosmos.esa.int/web/gaia/dpac/consortium}). Funding for the DPAC has been provided by national institutions, in particular the institutions participating in the {\it Gaia} Multilateral Agreement. This research made use of Lightkurve, a Python package for Kepler and TESS data analysis (Lightkurve Collaboration, 2018). This research has made use of NASA's Astrophysics Data System Bibliographic Services. 
H.N. acknowledges support from the European Research Council (ERC) under the European Union’s Horizon 2020 research and innovation programme (Grant Agreement No. 947660).

\section*{Data Availability}

 Observational data for analyzed targets are publicly available. Pulsation models will be shared upon request to the corresponding author.



\bibliographystyle{mnras}
\bibliography{references.bib} 





\bsp	
\label{lastpage}
\end{document}